\documentclass[notitlepage,superscriptaddress,twocolumn,tightenlines]{revtex4-2}

\usepackage[utf8]{inputenc}
\usepackage[english]{babel}
\usepackage{booktabs} 
\usepackage{comment} 
\usepackage{hyperref} 
\usepackage{physics} 
\usepackage[table,xcdraw,dvipsnames]{xcolor} 

\allowdisplaybreaks

\usepackage{lipsum}
\usepackage{xfrac}

 \usepackage{multirow}

\usepackage{amsmath,verbatim,latexsym,amssymb,indentfirst,mathrsfs,mathtools,amsthm,bbm,bm,hyperref,url,cancel,subcaption}
\usepackage[font=small,labelfont=bf,format=plain,justification=raggedright,singlelinecheck=false]{caption}
\usepackage{verbatim,indentfirst}
\usepackage[title,titletoc]{appendix}

\usepackage{graphicx}
\graphicspath{{images/}}
\usepackage{epstopdf}
\newcommand{\caphead}[1]{{\bf #1}}

\renewcommand{\thesection}{\Roman{section}}
\renewcommand{\thesubsection}{\Roman{section} \Alph{subsection}}
\renewcommand{\thesubsubsection}{\Roman{section} \Alph{subsection} \arabic{subsubsection}}
\makeatletter
\def\p@subsection{}
\makeatother
\makeatletter
\def\p@subsubsection{}
\makeatother

\makeatletter
\newcommand\footnoteref[1]{\protected@xdef\@thefnmark{\ref{#1}}\@footnotemark}
\makeatother

\usepackage{soul}

\usepackage{color}
\usepackage[dvipsnames]{xcolor}
\usepackage[normalem]{ulem}

\newcommand{\new}[1]{{\color{Black}#1}}

\newcommand{\Ent}{{\rm E}}

\newcommand{\Sites}{N}  

\newcommand{\tot}{ {\rm tot} }
\def\id{\mathbbm{1}}   

\newcommand{\LParen}{ \bm{(} }
\newcommand{\RParen}{ \bm{)} }
\newcommand{\JParen}{ {(j)} }

\renewcommand\th{ {\rm th} }

\begin{document}

\title{Non-Abelian symmetry can increase entanglement entropy}

\author{Shayan Majidy} 
\email{smajidy@uwaterloo.ca}
\affiliation{Institute for Quantum Computing, University of Waterloo, Waterloo, Ontario N2L 3G1, Canada}
\affiliation{Perimeter Institute for Theoretical Physics, Waterloo, Ontario N2L 2Y5, Canada} 

\author{Aleksander Lasek}
\affiliation{Joint Center for Quantum Information and Computer Science, NIST and University of Maryland, College Park, Maryland 20742, USA}

\author{David A. Huse}
\affiliation{Department of Physics, Princeton University, Princeton, New Jersey 08544, USA}

\author{Nicole Yunger Halpern}
\email{nicoleyh@umd.edu}
\affiliation{Joint Center for Quantum Information and Computer Science, NIST and University of Maryland, College Park, MD 20742, USA}
\affiliation{Institute for Physical Science and Technology, University of Maryland, College Park, MD 20742, USA}

\date{\today}

\begin{abstract} 
The pillars of quantum theory include entanglement and operators' failure to commute. The Page curve quantifies the bipartite entanglement of a many-body system in a random pure state. This entanglement is known to decrease if one constrains extensive observables that commute with each other (Abelian ``charges''). Non-Abelian charges, which fail to commute with each other, are of current interest in quantum thermodynamics. For example, noncommuting charges were shown to reduce entropy-production rates and may enhance finite-size deviations from eigenstate thermalization. Bridging quantum thermodynamics to many-body physics, we quantify the effects of charges' noncommutation---of a symmetry's non-Abelian nature---on Page curves. First, we construct two models that are closely analogous but differ in whether their charges commute. We show analytically and numerically that the noncommuting-charge case has more entanglement. Hence charges' noncommutation can promote entanglement.
\end{abstract}

\maketitle

\section{Introduction}

Entanglement has illuminated quantum many-body phenomena from space-time's structure~\cite{van2010building, swingle2018spacetime, cao2017space} to phases~\cite{skinner2019measurement, li2019measurement, chan2019unitary, somma2004nature, geraedts2017characterizing,  bao2020theory, gullans2020dynamical, wen2019choreographed, zeng2015quantum} 
and thermalization~\cite{abanin2019colloquium}. A large, isolated many-body system thermalizes internally when evolved under a nonintegrable, chaotic Hamiltonian. Such dynamics tend to imbue an initial pure state, after long times, with properties closely approximated in pure states drawn randomly from the available Hilbert space. The random state's average bipartite entanglement is quantified with a \emph{Page curve}~\cite{page1993average}: Consider partitioning the system into two subsystems, calculating a subsystem's entanglement entropy, and averaging the entropy over states drawn randomly from the full system's Hilbert space. The average, plotted against the subsystem's size, forms a Page curve.

Page curves have been studied in the context of Abelian symmetries~\cite{altland2022maximum, monteiro2021quantum, nakagawa2018universality, hackl2019average, hackl2021bosonic, vidmar2017entanglement, bianchi2019typical, bianchi2021volume, murciano2022symmetry, goldstein2018symmetry, xavier2018equipartition, calabrese2021symmetry, laflorencie2014spin, bonsignori2019symmetry, belin2013holographic, cornfeld2018imbalance, barghathi2018renyi, tan2020particle, murciano2020symmetry, kiefer2020evidence}. Consider a many-body system whose evolution conserves an extensive observable, or \emph{charge}; examples include the total particle number. Studying thermalization properties via a Page curve, one draws random pure states from a chosen particle-number sector---an eigenspace of the charge. We call such an eigenspace a \emph{microcanonical} subspace $\cal{S}$. More generally, the system may have multiple charges that commute with each other, so that the symmetry remains Abelian.  $\cal{S}$ can be chosen to be an eigenspace shared by the charges (apart from the Hamiltonian).

However, noncommutation lies at the heart of quantum theory, underlying uncertainty relations, measurement disturbance, and notions of locality~\cite{grimmer2021measurements, haag2012local}.
Conserved charges can fail to commute with each other,
though charges' commutation was assumed implicitly across thermodynamics for decades~\cite{balian1987equiprobability, jaynes1957information, yunger2018beyond, murthy2022non, yunger2016microcanonical, guryanova2016thermodynamics, lostaglio2017thermodynamic}.
The assumption was lifted in quantum thermodynamics recently~\cite{lostaglio2017thermodynamic, guryanova2016thermodynamics, yunger2018beyond, yunger2016microcanonical, sparaciari2018first, khanian2020from, khanian2020resource, manzano2022non, yunger2022build, kranzl2022experimental,  vaccaro2011information, gour2018Quantum, popescu2018quantum, popescu2019reference, ito2018Optimal, bera2019thermo, manzano2018Squeezed,  yunger2020noncommuting, fukai2020noncommutative,  mitsuhashi2021characterizing, croucher2018information, wright2018quantum, zhang2020stationary, medenjak2020isolated, croucher2021memory, marvian2021qudit, marvian2022rotationally, murthy2022non, ducuara2022quantum, hinds2018quantum, marvian2022restrictions, strasberg2022classicality}.
\new{Charges' noncommutation has
} 
been shown to conflict with derivations of the thermal state's form~\cite{yunger2018beyond,yunger2016microcanonical}; reduce entropy-production rates~\cite{manzano2018Squeezed}; and 
conflict with the eigenstate thermalization hypothesis, a framework for explaining quantum systems' internal thermalization~\cite{murthy2022non}. The experimental testing of these results~\cite{yunger2022build} has begun with a trapped-ion simulator~\cite{kranzl2022experimental} 
\new{whose dynamics were chaotic~\cite{yunger2020noncommuting}, yet conserved all three components of the global spin. 
}
Inspired by quantum thermodynamics, we aim to quantify a particularly quantum feature of many-body physics in this paper: how charges' noncommutation---a symmetry's non-Abelian nature---affects Page curves.

This comparison calls for two models that parallel each other closely, yet differ in whether their charges commute. Whether such models exist, what ``parallel closely'' should mean, and how to construct such models is unclear. We therefore posit criteria to encapsulate models' analogousness. Furthermore, we construct two models that meet these criteria. Each model consists of two-qubit sites. Every local charge is a product of two-qubit Pauli operators and/or identity operators.

We compare these models' Page curves in two settings. Conventional thermodynamics suggests one: a microcanonical subspace, a simultaneous eigenspace of the charges. The noncommuting-charge model has only one microcanonical subspace, because noncommutation tends to block observables from having well-defined values simultaneously. Also, the notion of a microcanonical subspace has been generalized to an \emph{approximate microcanonical (AMC) subspace}, to accommodate noncommuting charges~\cite{yunger2016microcanonical,yunger2020noncommuting,kranzl2022experimental}. Here, every charge has a fairly well-defined value: Measuring any charge has a high probability of yielding an outcome close to the expected value. We identify AMC subspaces in the noncommuting-charge model and analogs in the commuting-charge model. Each pair of such subspaces yields another pair of Page curves.

We estimate the Page curves numerically and, in the microcanonical comparison, analytically. In every setting where we can do so, the nonconcommuting-charge Page curve lies above the commuting-charge curve. On average, therefore, charges' noncommutation appears to promote entanglement. For systems of $N \gg 1$ sites, the Page curves' separation decreases, but only polynomially in the system size, as $1/N$. We posit that the gap arises solely from whether the charges commute, due to the close parallel between our two models. This conjecture calls for testing with more parallel models and for more-general explanations, which we \new{partially} leave as a challenge for future research. \new{Yet we find that, in the microcanonical comparison, the gap arises from state-counting effects---noncommuting charges' allowing the subspace to be larger than commuting charges do.
Furthermore, we posit an explanation based on each subspace's minimally entangled basis.
Our findings are suggestive of how charges' noncommutation affects quantum many-body phenomena such as thermalization.
}

The rest of this paper is organized as follows. 
\new{In Sec.~\ref{sec:Page}, we overview Page curves;
in Sec.~\ref{sec:analogous_models}, we present the analogous models. We compare the models' Page curves using microcanonical subspaces (Sec.~\ref{sec:s0_subspace}), 
then using AMC subspaces (Sec.~\ref{sec:AMC_subspace}). 
Section~\ref{sec:discussion} concludes with opportunities established by this work.
}

\section{Page-curve background} \label{sec:Page}
To introduce Page curves, we must introduce entanglement entropy.
Consider an isolated (``global'') system, associated with a Hilbert space $\cal{H}$, in a pure state $\ket{\Phi}$. Denote by $A$ a subsystem associated with a dimension-$D_A$ Hilbert space. Denote by $B$ the rest of the system.  The full system's Hilbert space is the outer product of the subsystems' Hilbert spaces. The \textit{entanglement entropy} is the von Neumann entropy of $\rho_A \coloneqq \Tr_{B}(\dyad{\Phi})$~\cite{nielsen2002quantum}:
\begin{equation}
    S_{\Ent} 
    \coloneqq S(\rho_A) \coloneqq -\Tr(\rho_A\log \rho_A)
    \leq \log D_A . \label{eq:EE_onestate}
\end{equation}
The logarithms are base $e$, giving entropies in units of nats. $A$ is entangled with $B$ if $S_{\Ent} > 0$. 

The Page curve quantifies the average entanglement in a subspace $\cal{S}$ of interest. Let $A$ consist of $N_A$ identical sites, and let $B$ consist of $N_B$ more, such that $N_A + N_B = N$.
Consider selecting a global pure state from $\cal{S}$ uniformly randomly---according to the Haar measure~\cite{penny2021understanding}. Calculating $S_\Ent$, then averaging over Haar-random states, yields
\begin{align}
   \label{eq:EE}
   \expval{S_{\Ent}}_{\cal{S}} &\coloneqq 
   - \expval{\Tr(\rho_A\log  \rho_A)}_{\cal{S}}. 
\end{align}
Plotted against $N_A$, $\expval{S_{\Ent}}_{\cal{S}}$
forms the Page curve for subspace $\cal{S}$~\cite{page1993average}.

We estimate the curve numerically as follows. 
Denote by $\{ \ket{\psi_{\ell} } \}$ any basis for the subspace.
We weight the $\ell^\th$ element with a random number $c_\ell$ drawn from a complex normal distribution. Summing the weighted elements, and renormalizing with a constant $C_{\text{norm}}$, we form a Haar-random state:
$\frac{1}{C_{\text{norm}}} \sum_\ell c_{\ell} \ket{\psi_{\ell}}$.
We sample $10^3$--$10^4$ states, calculate each state's $S_\Ent$, and average to estimate the Page curve.

In the best-known example, no charges constrain the system~\cite{page1993average}. Denote by $\cal{H}$ the full Hilbert space and by $d$ the local dimension (of a site's Hilbert space).
The unconstrained Page curve is, for $N_A\leq N_B$,
\begin{equation}
    \expval{S_E}_{\cal{H}} 
    \approx N_A \log d - \frac{1}{2} \, d^{N_A-N_B} . \label{eq:Page}
\end{equation}

The terms in Eq.~\eqref{eq:Page} stem from different physics, as do the analogous terms in constrained Page curves. Consider averaging the Haar-random states over $\cal{S}$ before calculating any entropy. The averaged state, $\langle \rho \rangle_{\cal{S}}$, is the maximally mixed state within $\cal{S}$. Tracing out $B$ yields $\expval{\rho_A}_{\cal{S}} \coloneqq \Tr_{B}( \expval{\rho}_{\cal{S}})$, whose entropy follows from state-counting arguments (App.~\ref{app:analytical}). We therefore call $S\left(\expval{\rho_A}_{\cal{S}}\right)$ the subspace-$\cal{S}$ Page curve's \textit{state-counting term}. In terms of it, the curve decomposes as
\begin{align}
    \expval{S_{\Ent}}_{\cal{S}} &= S\left(\expval{\rho_A}_{\cal{S}}\right)  + \left[\expval{S_{\Ent}}_{\cal{S}} -S\left(\expval{\rho_A}_{\cal{S}}\right)  \right]. \label{eq:splittingEE}
\end{align}
Since $\expval{\rho}_{\cal{S}}$ is maximally mixed, $S\left(\expval{\rho_A}_{\cal{S}}\right)$ equals the greatest possible entropy: $\expval{S_{\Ent}}_{\cal{S}} \leq S\left(\expval{\rho_A}_{\cal{S}}\right)$. Hence the bracketed term in Eq.~\eqref{eq:splittingEE} is $\leq 0$. That term encodes the interference between different states' contributions to the Page curve's Haar average. This \textit{interference term} is exponentially small in $N_B - N_A$ \cite{page1993average}. In the unconstrained curve~\eqref{eq:Page}, $N_A\log d$ is the state-counting term, and $- \frac{1}{2} d^{N_A-N_B}$ is the interference term. 

\section{Analogous noncommuting-charge and commuting-charge models} \label{sec:analogous_models}

We aim to identify how charges' noncommutation affects the Page curve. Therefore, we need two models that differ in whether their charges commute and otherwise differ minimally. Whether such models exist, what ``differ minimally'' should mean, and how to construct such models is unclear. For instance, the most commonly studied non-Abelian symmetry group is SU(2); the associated charges are the Pauli operators, $X$, $Y$, and $Z$. How to construct an analogous model with three commuting charges is not obvious. For example, the group U(1)$^{\times 3}$ is generated by three charges that commute but are not multiplicatively interrelated. In contrast, $X Y = iZ$.  

We address this challenge by positing five criteria that capture what renders noncommuting-charge and commuting-charge models analogous. Then, we construct two models that meet these criteria.  We denote by $Q_\alpha^\tot$ the global noncommuting charges and by $C_{\alpha}^\tot$ the global commuting charges. The criteria concern also the subspaces used to calculate the Page curves. Denote by $\ket{\psi}$ any state from the noncommuting-charge subspace $\cal{N}$. Measuring $Q_\alpha^\tot$ yields outcome $\gamma$ with some probability. This probability, averaged over the $\ket{\psi}$, we denote by $p_\alpha^{\cal{N}}(\gamma)$. Define $p_{\alpha}^{\cal{C}}(\gamma)$ analogously for the commuting-charge subspace $\cal{C}$.

We define as analogous any commuting-charge and noncommuting-charge models that satisfy five criteria:
\begin{enumerate}
    \item In each model, the system consists of $N$ sites, 
    each formed from a $d$-level qudit. Each model has $c$ constrained global charges. \label{crit1} 
    
    \item Each global charge (i) is a sum of single-site observables and (ii) acts nontrivially and identically on all sites.\label{crit2}
    
    \item Each charge $Q_\alpha^\tot$ has the same spectrum as its analog $C_{\alpha}^\tot$.
    \label{crit3}
    
    \item \new{Any two commuting charges form a product analogous to the analogous noncommuting charges' product.}
    \label{crit4}
    
    \item The constrained subspaces, $\cal{N}$ and $\cal{C}$, 
    are such that $p_\alpha^{\cal{N}}(\gamma) = p_\alpha^{\cal{C}}(\gamma)$.
    \label{crit5}
\end{enumerate}

\begin{figure}
    \centering
    \includegraphics[width=0.85\columnwidth]{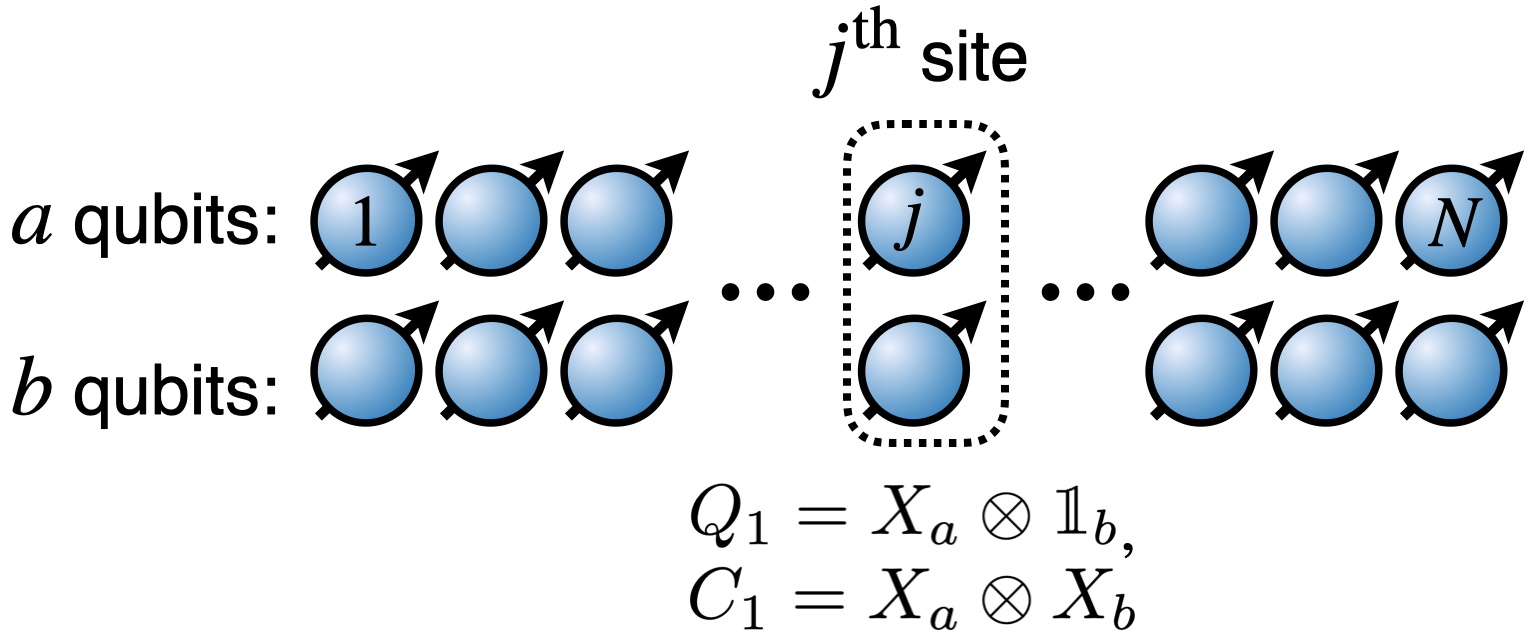}
    \caption{\caphead{Analogous noncommuting-charge and commuting-charge models.} Each model consists of $N$ sites. A site consists of two qubits, $a$ and $b$. The local noncommuting observables of interest include $Q_1$; and the local commuting observables, $C_1$.
    }
    \label{fig:schematic}
\end{figure}

We now construct two models that satisfy the criteria (Fig.~\ref{fig:schematic}). 
Each global charge ($Q_\alpha^\tot$ or $C_\alpha^\tot$) follows from summing single-site observables $Q_\alpha$ or $C_\alpha$. Denote by $Q_\alpha^{(j)}$ an observable defined on site $j$'s Hilbert space, and define $C_\alpha^{(j)}$ analogously. The global charges are extensive: If $\id$ denotes the single-site identity operator,
\begin{align}
   Q_\alpha^\tot  
   \coloneqq \sum_{j = 1}^\Sites \id^{\otimes (j - 1)}  \otimes Q_\alpha^\JParen  \otimes 
   \id^{\otimes (\Sites - j) }
   \equiv   \sum_{j = 1}^{\Sites}  Q_\alpha^\JParen ,
   \label{eq:globalcharge}
\end{align}
and $C_{\alpha}^\tot 
\coloneqq \sum_{j = 1}^{\Sites}  C_\alpha^\JParen$. 

The noncommuting charges can generate $\mathfrak{su}(2)$ if each site contains one qubit ($d=2$). By criterion~\ref{crit2}, three charges impose three restrictions on each site. A fourth restriction follows from the normalization of the site's state. These restrictions suggest that, to support a model with three commuting charges, $d$ should be $\geq 4$. Choosing $d=4$ for simplicity, we form each site's qudit from two qubits, $a$ and $b$. The noncommuting local observables are
\begin{equation}
    Q_1 =  X_a \otimes \id_b, \;\; 
    Q_2 =  Y_a \otimes \id_b, \; \text{and} \; 
    Q_3 =  Z_a \otimes \id_b;
    \label{eq:NC_charges}
\end{equation}
and the commuting local observables are
\begin{equation}
    C_1 = X_a \otimes X_b, \;\; 
    C_2 = Y_a \otimes Y_b, \; \text{and} \; 
    C_3 = Z_a \otimes Z_b. \label{eq:CC_charges}
\end{equation}

These models satisfy criteria \ref{crit1}--\ref{crit3} overtly and by simple calculation. Criterion \ref{crit4} concerns \new{products of charges. 
}
For unequal indices $\alpha, \beta, \gamma \in \{1,2,3\}$,
\begin{align}
    Q_{\alpha} Q_\beta = i \epsilon_{\alpha\beta\gamma} Q_\gamma,
   \quad \text{and} \quad 
    C_{\alpha} C_\beta = - C_\gamma .
\end{align}
These equations parallel each other because multiplying two distinct charges yields the third charge times a constant.
Furthermore, $Q_\alpha Q_\alpha = C_\alpha C_\alpha = \id \; \, \forall \alpha$. 

Criterion~\ref{crit5} is satisfied if we choose subspaces adroitly. In the microcanonical subspaces identified below, the $p_\alpha^{\cal{N}}(\gamma)$'s and $p_\alpha^{\cal{C}}(\gamma)$'s equal Kronecker delta functions and so each other. As detailed below, we can also construct AMC subspaces such that $p_\alpha^{\cal{N}}(\gamma) = p_\alpha^{\cal{C}}(\gamma)$ for all $\alpha$ and $\gamma$.

\section{Microcanonical-subspace comparison}\label{sec:s0_subspace}

The noncommuting-charge model has exactly one microcanonical subspace, ${\cal{N}}_0$: the eigenvalue-0 eigenspace shared by $Q^\tot_{1,2,3}$. This subspace exists only if $N$ is even. The analogous commuting-charge subspace, ${\cal{C}}_0$, 
is the eigenvalue-0 eigenspace shared by $C_{1,2,3}^\tot$.
This subspace exists only if $N$ equals a multiple of four (App.~\ref{app:commuting_model}).


\begin{figure}
     \centering
     \begin{subfigure}[b]{.9\columnwidth}
         \centering
         \includegraphics[width=\columnwidth]{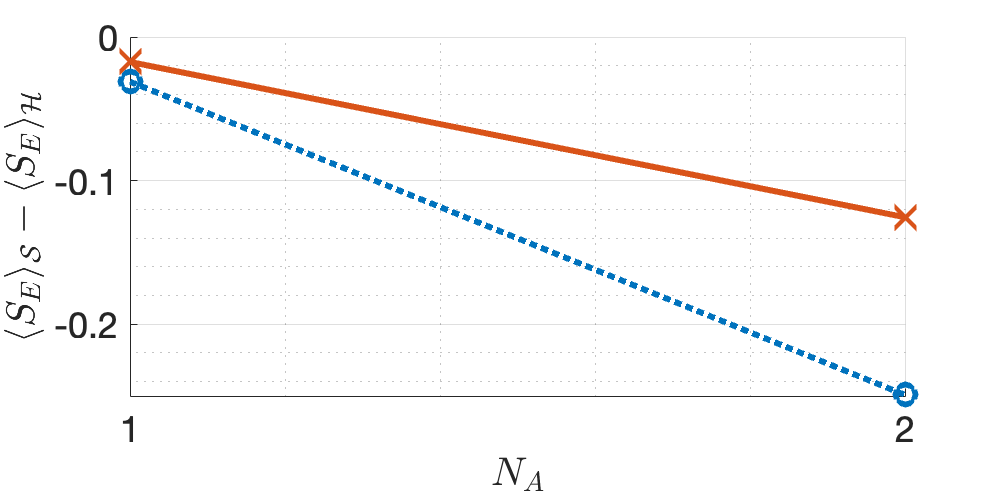}
         \caption{Global-system size $N=4$}
     \end{subfigure}
     \hfill
     \begin{subfigure}[b]{.9\columnwidth}
         \centering
         \includegraphics[width=\columnwidth]{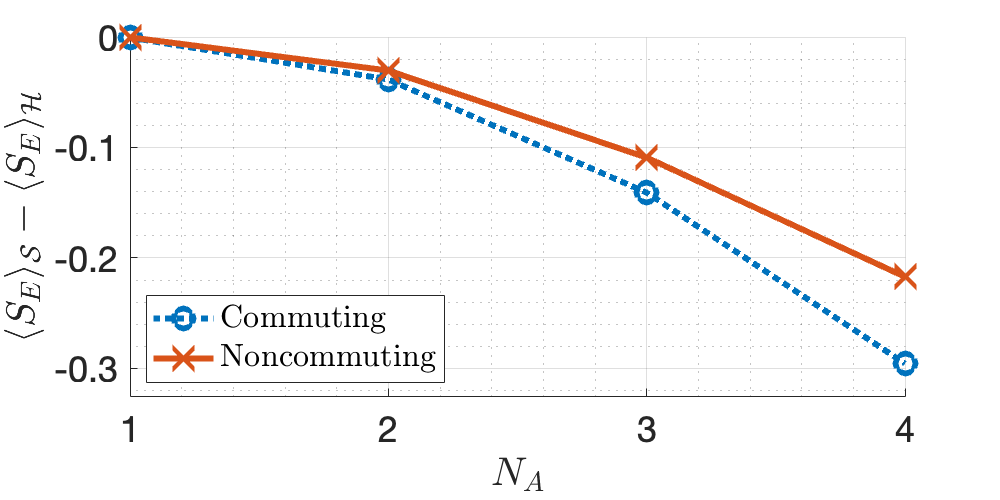}
         \caption{Global-system size $N=8$}
     \end{subfigure}
        \caption{\caphead{Page curves constructed from microcanonical subspaces.} $ \expval{S_{\Ent}}_{\cal{S}}$ denotes any Page curve restricted by charges; and  $ \expval{S_E}_{\cal{H}}$, the unrestricted Page curve. 
        The red x's form the noncommuting-charge model's Page curve, and the circular blue markers form the commuting-charge model's Page curve. The connecting lines guide the eye. We calculated the top panel's ($N=4$) Page curves from $10^4$ samples each and the bottom panel's ($N=8$) Page curves from $10^3$ samples each.
        The $x$-axis ends at $N_A = N/2$ for conciseness; the Page curves are symmetric according to numerics.
        }
    \label{fig:s0_change_EE}
\end{figure}

We estimate Page curves numerically using the procedure outlined in \new{Sec.~\ref{sec:Page} and using~\cite{qetlab}.
}
Here, $\expval{S_{\Ent}}_{\cal{S}}$ denotes 
the Page curve for a subspace $\cal{S}$, and $\expval{S_E}_{\cal{H}}$ denotes 
the unrestricted Page curve~\eqref{eq:Page}. To highlight the gap between the noncommuting-charge and commuting-charge curves, we plot
$\expval{S_{\Ent}}_{\cal{S}} - \expval{S_E}_{\cal{H}}$ for ${\cal{S}=\cal{N}}_0$, ${\cal{C}}_0$ in Fig.~\ref{fig:s0_change_EE}. 
At all partition locations $N_A$, the noncommuting-charge Page curve lies above the commuting-charge Page curve. 
For example, at the midpoint ($N_A = N/2$), the gap is $0.124$ nats ($17.8 \%$ of the average of the two Page curves at $N_A = N/2$) when $N=4$ and $0.0797$ nats ($10.5 \%$) when $N=8$. In this microcanonical case, therefore, the subspace constrained by noncommuting charges has more entanglement, on average. 

We posit the following explanations for this phenomenon in our setting. 
\new{First, the subspace's dimensionality upper-bounds the entanglement entropy:  $S_\Ent \leq \log D_A$
[Eq.~\eqref{eq:EE_onestate}].
The bound tends to approximate random states' entropies.
Hence one might expect a higher Page curve of whichever model has the larger subspace. Indeed, the noncommuting-charge subspace is of dimensionality 32, when $\Sites=4$, exceeding the commuting-charge dimensionality of 24. 
When $\Sites = 8$, the noncommuting-charge dimensionality is 3584, again exceeding its commuting-charge analog, 2520.
Our analytical results below agree at large $N$: 
The noncommuting-charge curve lies above the commuting-charge curve if approximated with the state-counting term, which depends essentially on subspace dimensionality. 

We expect this dimensionality argument to explain our results only partially, the Page curves do not saturate the upper bound~\eqref{eq:EE_onestate}.
Hence we posit that, when the compared subspaces have similar dimensionalities, their minimally entangled bases may help determine the Page curves' relative locations.
}
The commuting-charge model's microcanonical subspace, $\mathcal{C}_0$, has a tensor-product basis. 
The reason is, every global charge $C_{\alpha}^\tot$ commutes with
all the subsystem charges $C_{\alpha'}^{(A)}$ and $C_{\alpha''}^{(B)}$.
In contrast, in the noncommuting-charge model,
each global charge $Q_\alpha^\tot$ fails to commute with some subsystem charges $Q_{\alpha'}^{(A)}$ and $Q_{\alpha''}^{(B)}$.
Hence the microcanonical subspace $\mathcal{N}_0$ has no tensor-product basis. Therefore, the minimally entangled basis has more entanglement in the noncommuting-charge model. Hence one might expect a higher Page curve there. \new{Additionally, in App.~\ref{app:additivity_ansatz}, we show that sequentially introducing charges restricts the Page curve subadditively if the charges fail to commute, and superadditively if the charges commute, at finite $N$.
}


We now analytically calculate the difference between the noncommuting-charge and commuting-charge Page curves in these microcanonical subspaces at large $N$.
Recall that the interference term [Eq.~\eqref{eq:splittingEE}] is exponentially small in $N_B - N_A$ \cite{page1993average}. Consequently, the state-counting term approximates the Page curve when $N_B -N_A \gg 1$~\footnote{Computational restrictions prevent $N_B - N_A$ from growing very large in the numerics. Therefore, we refrain from plotting our analytics in Fig.~\ref{fig:s0_change_EE}.}. We calculate state-counting terms  in App.~\ref{app:analytical}, using large-$N$ expansions. 
We assume that $N_A, N_B = \mathcal{O}(N)$; the subsystems' sizes are near their average values.
Both subspaces' Page curves have the leading, $\mathcal{O}(N^0)$ term
\begin{equation}
    L \coloneqq N_A\log d - \frac{3}{2}\log \frac{N}{N_B} + \frac{3N_A}{2N} \, .
\end{equation}
The noncommuting-charge Page curve is
\begin{align}
     L + \frac{3N_A}{4 N^2} + \frac{N_A^2}{2 N^2 N_B} +  \mathcal{O}\left(N^{-\frac{3}{2}}\right) \, , \label{eq:NC_pagecurve}
\end{align}
and the commuting-charge Page curve is~\footnote{In both expressions, the 
$\mathcal{O}\left(N^{-\frac{3}{2}}\right)$ term may vanish, so the next nonzero term may be
$\mathcal{O}\left(N^{-2}\right)$.}
\begin{align}
    L + \frac{3N_A}{4 N^2} - \frac{N_A^2}{2 N^2 N_B} +  \mathcal{O}\left(N^{-\frac{3}{2}}\right). \label{eq:CC_pagecurve}
\end{align}
The noncommuting-charge Page curve is greater by an amount $\frac{N_A^2}{N^2N_B}$, at leading order. The difference decreases as $N$ grows. 
This decline is consistent with the correspondence principle~\cite{shankar2012principles}---as systems grow large, they grow classical. Noncommutation is nonclassical, so its effects on observable phenomena should diminish as $N \to \infty$~\cite{yunger2016microcanonical}. More precisely, the charge densities 
$Q_\alpha^{\tot} / N$ have commutators that vanish in the thermodynamic limit~\cite{ogata2013approximating, yunger2016microcanonical}:
$[Q_\alpha^\tot / N, \; Q_{\alpha'}^\tot / N] \to 0$,
for all $\alpha$ and $\alpha'$, as $N \to \infty$. 
However, the Page-curve difference shrinks relatively slowly---as $1/N$, rather than exponentially---as $N$ grows.

\section{Approximate-microcanonical-subspace comparison}\label{sec:AMC_subspace}

Having compared our two models using microcanonical subspaces, we progress to AMC subspaces, generalizations that accommodate charge noncommutation~\cite{yunger2016microcanonical, kranzl2022experimental, yunger2020noncommuting}. Instead of having well-defined values in an AMC subspace, the charges have fairly well-defined values: Measuring any $Q_\alpha^\tot$ has a high probability of yielding an outcome close to the expected value.
This section outlines how to construct analogous AMC subspaces in the noncommuting-charge and commuting-charge models. We then compare the models' Page curves numerically. 
The noncommuting-charge Page curve is always higher, as in the microcanonical-subspace comparison.


One can construct as follows AMC subspaces in the noncommuting-charge model. Define the $a$-qubit magnetization $Z_a^\tot \coloneqq \sum_{j=1}^N Z_a$, which has eigenvalues $2m$. Define the $a$-type spin-squared operator $(\vec{S}_a^\tot)^2 
\coloneqq \sum_{a=1}^3  ( Q_a^\tot )^2$,
which has eigenvalues $s(s+1)$ (we set $\hbar = 1$).
$(\vec{S}_a^\tot)^2$ shares with $Z_a^\tot$---and so 
$(\vec{S}_a^\tot)^2 \otimes \id_b^\tot$ shares with $Q_3^\tot$---eigenspaces 
$\cal{N}$ labeled by the quantum numbers $(s,m)$. Some such eigenspaces are AMC subspaces, we find by direct calculation. For each $(s, m)$ value, we calculate the probability distributions $p_\alpha^{\cal{N}}(\gamma)$. Each distribution exhibits one peak, as required by the definition of ``AMC subspace,'' for certain $(s, m)$ (App.~\ref{app:eg_amc_probs}).
Having identified AMC subspaces defined by noncommuting charges, we construct analogs $\cal{C}$ defined by commuting charges. Appendix~\ref{app:eg_amc_probs} details the process. We identify six pairs of parallel (commuting-charge and noncommuting-charge) AMC subspaces, labeled by $s=m=1,N/2$, for $N=4,8$, as well as by $s=m=N/2$, for $N=2,6$.


We estimate each AMC subspace's Page curve numerically. In every comparison, the noncommuting-charge ($\cal{N}$) Page curve lies above its commuting-charge ($\cal{C}$) partner. An illustrative example is parameterized by $N=8$ and $s=m=1$. We compare the two curves at the midpoint $N_A = N/2$. Recall that $\expval{S_{\Ent}}_{\cal{S}}$ denotes a Page curve for the subspace $\cal{S}$. When $N_A = 4$, $\expval{S_{\Ent}}_{\cal{N}}- \expval{S_E}_{\cal{C}} = 0.027$ nats, which is $7.11 \%$ of the two Page curves' average. The percent difference varies across the AMC-subspace pairs from $0.272 \%$ to $7.11 \%$.
Hence charges' noncommutation increases the average entanglement entropy in AMC subspaces as in the microcanonical comparison.

\section{Outlook} \label{sec:discussion}

We have demonstrated that constrained charges' noncommutation promotes average entanglement. Numerical and analytical calculations support this conclusion in microcanonical and AMC subspaces. 
\new{In the microcanonical comparison, the Page-curve gap stems from the discrepancy between the subspaces' dimensionalities.
}
This work reveals how one hallmark of quantum theory---operators' failure to commute---influences another---entanglement. \new{Due to entanglement's role in thermalization, our results are suggestive of how charges' noncommutation affects quantum many-body thermalization (as discussed more below).}

Our conclusions rest on two models that resemble each other closely but differ in whether their charges commute. Our models can now be used to explore effects of charges' noncommutation on other quantum phenomena. Possibilities include chaos~\cite{stockmann2000quantum, D'Alessio_16_From}, as analyzed with out-of-time-ordered correlators \cite{hashimoto2017out, rozenbaum2017lyapunov, garcia2018chaos, alonso2022diagnosing} and random unitary circuits~\cite{fisher2022random, hulse2021qudit};
bounds on quantum-simulation errors~\cite{buluta2009quantum}; and quantum-machine-learning algorithms' performances~\cite{biamonte2017quantum}.

Additionally, our results raise a puzzle. We find that charges' noncommutation promotes entanglement, which accompanies thermalization. Another result links noncommuting charges to enhanced thermalization: Non-Abelian symmetries destabilize many-body localization, a phase of matter in which entanglement spreads very slowly~\cite{potter2016symmetry}. In contrast, charges' noncommutation was found to restrict thermalizing behaviors in two ways. First, local operators' time-averaged expectation values may deviate from thermal predictions by anomalously large corrections if charges fail to commute~\cite{murthy2022non}. Second, charges' noncommutation can decrease the rate of entropy production, which accompanies thermalization~\cite{manzano2022non}. These two results technically do not conflict with ours or with Ref.~\cite{potter2016symmetry}, concerning different setups.  However, they invite a more general understanding of when non-Abelian symmetries enhance or suppress entanglement and thermalization.


Apart from the foregoing theoretical opportunities, the difference between commuting-charge and noncommuting-charge entanglement entropies may be observed experimentally. For example, at the Page curves' midpoints ($N_A = N/2$), the difference is $0.124$ nats in the microcanonical setting for $N=4$. A precision of $\approx 0.05$ nats should therefore suffice to observe the difference. Such a precision has been achieved with trapped ions~\cite{linke2018measuring, islam2015measuring, brydges2019probing} and ultracold atoms~\cite{kaufman2016quantum, lukin2019probing, leonard2020signatures}. 
Furthermore, noncommuting-charge thermodynamics has been argued and demonstrated to be observable on these platforms~\cite{yunger2020noncommuting, yunger2022build, kranzl2022experimental}.

\begin{acknowledgments}
We thank Alex Altland, Hubert de Guise, Rajibul Islam, Nathaniel Johnston, Eduardo Martin-Martinez, Tobias Micklitz and Iosif Pinelis for helpful discussions and/or collaborations. 
This work received support from the National Science Foundation (QLCI grant OMA-2120757), the John Templeton Foundation (Award No. 62422), 
and the Vanier C.G.S. 
N.Y.H. thanks Princeton, and S.M. thanks QuICS, for their hospitality during the formation of this paper
\end{acknowledgments}

\begin{appendices}

\onecolumngrid
\renewcommand{\thesection}{\Alph{section}}
\renewcommand{\thesubsection}{\Alph{section} \arabic{subsection}}
\renewcommand{\thesubsubsection}{\Alph{section} \arabic{subsection} \roman{subsubsection}}
\makeatletter\@addtoreset{equation}{section}
\def\theequation{\thesection\arabic{equation}}

\section{Analytic expressions for state-counting terms in \\ microcanonical subspaces' Page curves} \label{app:analytical}

The Page curve \eqref{eq:EE} naturally splits into two terms, the state-counting term [$S(\expval{\rho_A}_{\cal{S}}) $ from Eq.~\eqref{eq:splittingEE}] and the interference term. The interference term is exponentially small in $N_B - N_A$. Thus, if $N_A \ll N_B$, the Page curve approximately equals the state-counting term. 
As explained in Sec.~\ref{sec:Page},
the state-counting term is easier to calculate than the Page curve is. We calculate the term in this appendix.

To recall the term's definition, consider a system restricted to a subspace $\cal{S}$ (e.g., a microcanonical or an AMC subspace) of dimensionality $D$. Denote by $\{ \ket{\psi_{\ell} } \}$ any orthonormal basis for the subspace. 
Taking any pure state from that subspace and Haar-averaging it
yields the maximally mixed state,
$\expval{\rho}_{\cal{S}} = \frac{1}{D} \sum_{\ell}^{D} \dyad{\psi_{\ell}}$.
Tracing out $B$ yields 
$\expval{\rho_A}_{\cal{S}} 
\coloneqq \Tr_{B}( \expval{\rho}_{\cal{S}})$,
whose entropy is the state-counting term:
\begin{equation}
    S(\expval{\rho_A}_{\cal{S}})  = - \Tr(\expval{\rho_A}_{\cal{S}} \log \expval{\rho_A}_{\cal{S}}). \label{eq:APP_STC}
\end{equation}
We calculate this term for microcanonical subspaces below.
First, we introduce notation, a technical tool, and assumptions (App.~\ref{app_Preliminaries}). We address the commuting-charge model in App.~\ref{app:commuting_model} and the noncommuting-charge model in App.~\ref{app:noncommuting_model}.

\subsection{Preliminaries}
\label{app_Preliminaries}

We use the following notation throughout this appendix. Denote by $X_a^\tot \coloneqq \sum_{j=1}^N  X_a$ the sum of the $a$ qubits' $X$ operators, and define $Y_a^\tot$ and $Z_a^\tot$ analogously. The $a$ qubits' total-spin-squared operator, $\vec{S}_a^2 = [(X_a^\tot)^2 + (Y_a^\tot)^2 + (Z_a^\tot)^2]/4$, has eigenvalues $s(s+1)$ (we set $\hbar = 1$). Denote by $m$ the $Z_a^\tot/2$ eigenvalue. Denote by $s_A$ subsystem $A$'s spin quantum number, and denote by $m_A$ subsystem $A$'s magnetic spin quantum number. Define $s_B$ and $m_B$ analogously. 

We will use Catalan's triangle, a triangular array of numbers related to the dimensionalities of qubit systems' Hilbert spaces~\cite{cohen2016entanglement, stanley2011enumerative}. The element in row $a$ and column $b$ is
\begin{equation}
    C_{a,b} = \frac{a-b+1}{a+1} {a+b \choose b}
    \, , \quad \text{for} \quad a \geq b \label{eq:Catalan}.
\end{equation}
The bound $a\geq b$ lends the array its triangular shape.
Temporarily consider an $N$-qubit system that has
quantum numbers $s$ and $m$.
For arbitrary $m$, $C_{\frac{N}{2}+s,\frac{N}{2}-s}$ equals the $s$ eigenspace's dimensionality. 

Throughout our approximations, we assume that parameters approximately equal their typical values: $m, s, m_A, s_A, m_B, s_B = \mathcal{O}\left(N^{-1/2}\right)$; and
$N_A, N_B = \mathcal{O}\left(N\right)$. 
We assume also that the global system is large: $N \gg 1$.

\subsection{Commuting-charge model's state-counting term} \label{app:commuting_model}

Appendix~\ref{app_CommQ_Constraints} describes how the commuting-charge model is constrained in a microcanonical subspace.
In App.~\ref{app:exact_4}, we calculate the commuting-charge state-counting term exactly. How the exact formula scales with $N$ is unclear. Therefore, we approximate the term to $\mathcal{O}(N^{-1})$ in App.~\ref{app:closed_6}, to identify differences from the noncommuting-charge model. 

\subsubsection{Constraints on commuting-charge model in microcanonical subspace} \label{app_CommQ_Constraints}

The microcanonical subspace $\mathcal{C}_0$ parallels the noncommuting-charge model's $s=m=0$ subspace. Let us specify quantitatively how the commuting-charge model is constrained. First, we introduce notation.

The local charges $C_{1,2,3}$ share four eigenstates,
the maximally entangled \emph{Bell states}~\cite{nielsen2002quantum}.
They are, if $\ket{\uparrow}$ and $\ket{\downarrow}$ denote the $Z$ eigenstates,
\begin{align}
    & \ket{\mathcal{B}_1} \coloneqq \frac{1}{\sqrt{2}}\left(
    \ket{\downarrow}_a\ket{\uparrow}_b - \ket{\uparrow}_a\ket{\downarrow}_b \right), \quad
    \ket{\mathcal{B}_2} \coloneqq \frac{1}{\sqrt{2}}\left(
    \ket{\downarrow}_a\ket{\downarrow}_b - \ket{\uparrow}_a\ket{\uparrow}_b \right), \\
    & \ket{\mathcal{B}_3} \coloneqq \frac{1}{\sqrt{2}}\left(
    \ket{\downarrow}_a\ket{\downarrow}_b + \ket{\uparrow}_a\ket{\uparrow}_b \right), 
    \quad \text{and} \quad
    \ket{\mathcal{B}_4} \coloneqq \frac{1}{\sqrt{2}}\left(
    \ket{\downarrow}_a\ket{\uparrow}_b + \ket{\uparrow}_a\ket{\downarrow}_b \right).
\end{align}
The Bell states correspond to the $(C_1,C_2,C_3)$ eigenvalues 
$(-1,-1,-1)$, $(-1,1,1)$, $(1,-1,1)$, and $(1,1,-1)$, respectively. We will use a $\mathcal{C}_0$ basis formed from tensor products of single-site Bell states. For a given basis state, let $P_k$ denote the number of sites in Bell state $k$.

Having specified notation, we use it to derive constraints on the system. The microcanonical subspace ${\cal{C}}_0$ is the eigenvalue-0 eigenspace of $C_{1,2,3}^\tot$, by analogy with the noncommuting-charge $s=0$ subspace.
If the global system is in an eigenvalue-0 eigenstate of $C_{1}^{\tot}$, then $P_1 + P_2 = P_3 + P_4 = \frac{N}{2}$. 
If the system is in an eigenvalue-0 eigenstate of $C_{2}^{\tot}$, then $P_1 + P_3 = P_2 + P_4  = \frac{N}{2}$. If the system is in an eigenvalue-0 eigenstate of $C_{3}^{\tot}$, then $P_1 + P_4 = P_2 + P_3  = \frac{N}{2}$. Together, these constraints imply 
\begin{align}
   \label{eq_Pop_Spread}
   P_1 = P_2 = P_3 = P_4 = N/4 \, .
\end{align} 
Since $N$ is an integer, these constraints can be met if $N$ is a multiple of $4$, which we assume.

\subsubsection{Exact expression for the commuting-charge state-counting term}\label{app:exact_4}

We first calculate $\expval{\rho_A}_{{\cal{C}}_0}$,
the reduced state of system $A$ when the global system is maximally mixed. In addition to the definitions above, we invoke the ``quadnomial" coefficient  
${n \choose k_1, k_2, k_3, k_4} \coloneqq \frac{n!}{k_1!k_2!k_3!k_4!}$.
Under the population restriction~\eqref{eq_Pop_Spread}, the global system's Hilbert space is of dimensionality
\begin{align}
   D = {N \choose \frac{N}{4},\frac{N}{4},\frac{N}{4},
    \frac{N}{4}} .
\end{align}
Denote by $A_{k}$ the number of $A$ sites in the Bell state $\ket{\mathcal{B}_k}$, and denote by $B_k$ the number of $B$ sites in $\ket{\mathcal{B}_k}$.
The global system is restricted to a subspace of dimensionality
\begin{align}
    D_A &= {N_A \choose A_1, A_2,A_3,A_4} {N_B \choose B_1,B_2,B_3,B_4} \\ &= {N_A \choose \frac{N_A}{4} + m_1, \frac{N_A}{4} +  m_2, \frac{N_A}{4} +  m_3, \frac{N_A}{4} + m_4} 
    {N_B \choose \frac{N_B}{4} - m_1, \frac{N_B}{4} -  m_2, \frac{N_B}{4} -  m_3, \frac{N_B}{4} - m_4}.
\end{align}
In accordance with Eq.~\eqref{eq_Pop_Spread}, $A_k+B_k=N/4$.
Furthermore, $A$ is restricted to a subspace of dimensionality
\begin{align}
    d_A &= {N_A \choose A_1, A_2,A_3,A_4} = {N_A \choose \frac{N_A}{4} + m_1, \frac{N_A}{4} +  m_2, \frac{N_A}{4} +  m_3, \frac{N_A}{4} + m_4}.
\end{align}

The global maximally mixed state is
$\expval{\rho}_{{\cal{C}}_0} = \frac{1}{D} \sum_{\ell=1}^{D} \dyad{\psi_{\ell}}$; the sum runs over all the states in our basis for $\mathcal{C}_0$. Denote by $\{\ket{A_1,A_2,A_3,A_4,i}\}$ a basis for subsystem $A$'s Hilbert space. The index $i$ distinguishes basis states that share the same $A_1,A_2,A_3,$ and $A_4$. Tracing out subsystem $B$ yields
\begin{equation}
     \expval{\rho_A}_{{\cal{C}}_0} = \frac{1}{D} \sum_{A_1,A_2,A_3,A_4,i} \frac{D_A}{d_A} \dyad{A_1,A_2,A_3,A_4,i}.
\end{equation}
The $\frac{D_A}{d_A}$ equals the dimensionality of the subsystem-$B$ subspace that is consistent with the subsystem-$A$ populations $A_1,A_2,A_3,$ and $A_4$. Taking the spectral decomposition, we calculate $ \expval{\rho_A}_{{\cal{C}}_0}$'s entropy and so the state-counting term:
\begin{align}
     S(\expval{\rho_A}_{{\cal{C}}_0}) &=  - \sum_{A_1, A_2, A_3, A_4} \frac{D_A}{D} \log(\frac{D_A}{ d_AD}) \label{app:eq_sub_in}\\
    &=  - \sum_{A_1,A_2,A_3,A_4} {N_A \choose A_1, A_2, A_3, A_4}
     \frac{{N_B \choose B_1,B_2,B_3,B_4}}{{N \choose \frac{N}{4},\frac{N}{4}, \frac{N}{4}, \frac{N}{4}
    }} \log\left( \frac{{N_B \choose B_1,B_2,B_3,B_4}}{{N \choose \frac{N}{4},\frac{N}{4},\frac{N}{4},\frac{N}{4}}} 
     \right).
\end{align}

\subsubsection{Closed-form approximation to the commuting-charge state-counting term}\label{app:closed_6}

Let us approximate the $\frac{D_A}{D}$ in Eq.~\eqref{app:eq_sub_in} as a Gaussian function. Via differentiation, we determine that $\log(\frac{D_A}{D})$ maximizes at $m_k = 0$ for all $k$. We Taylor-expand $\log(\frac{D_A}{D})$ around this maximum, keeping only terms larger than $\mathcal{O}\left(N^{-3/2}\right)$.
For conciseness, we define 
$c \coloneqq \frac{2N}{N_AN_B} = O \left( \frac{1}{N} \right)$, 
$d \coloneqq \frac{1}{3} \left(\frac{8}{N_B^2} -  \frac{8}{N_A^2}\right) 
= O \left( \frac{1}{N^2} \right)$, 
$f \coloneqq \frac{1}{2} 
\left(\frac{8}{N_A^2} +  \frac{8}{N_B^2}\right) 
= O \left( \frac{1}{N^2} \right)$, and 
$g \coloneqq \frac{1}{2} 
\left(\frac{32}{3N_A^3} +  \frac{32}{3N_B^3}\right) 
= O \left( \frac{1}{N^3} \right)$. 
We substitute these definitions into the expansion of $\log(\frac{D_A}{D})$:
\begin{align}
    \log(\frac{D_A}{D}) =& \; \log \left(\frac{2 c^{3/2}}{\pi ^{3/2}}\right) -c \left( \sum_{i=1}^{4}m_i^2 \right) -d \left(\sum_{i=1}^{4}m_i^3\right) +f \left(\sum_{i=1}^{4}m_i^2 \right) -g \left(\sum_{i=1}^{4}m_i^4 \right) +\frac{5}{4 N}-\frac{5}{4 N_A}  - \frac{5}{4 N_B} \nonumber \\ & + \mathcal{O}\left(N^{-3/2}\right).
\end{align}
Exponentiating each side yields
\begin{align}
   \frac{D_A}{D} =& \;  2\left(\frac{c}{ \pi}\right)^{\frac{3}{2}} \exp(- c \sum_{i=1}^4 m_i^2) \Bigg[ 1 - d \left(\sum_{i=1}^{4}m_i^3\right) + \frac{d^2}{2} \left( \sum_{i=1}^{4}m_i^3  \right)^2  + f \left(\sum_{i=1}^{4}m_i^2 \right) -g \left(\sum_{i=1}^{4}m_i^4 \right) +\frac{5}{4 N} -\frac{5}{4 N_A} \nonumber \\ & -\frac{5}{4 N_B} + \mathcal{O}\left(N^{-3/2}\right) \Bigg]. \label{app:Da/D_approx_2}
\end{align}
We check that this function is normalized to $O( N^{-3/2 } )$ (as $\frac{D_A}{D}$ should be normalized), but omit the check from this appendix.

Having approximated the first factor in the state-counting term~\eqref{app:eq_sub_in}, we address the second, 
$\log \left( \frac{D_A}{d_A D} \right)$.
By Stirling's approximation~\eqref{eq_Stirling},
\begin{align}
   \label{eq_Stirling}
   \log(n) = n\log(n) - n + \frac{1}{2}\log(2\pi n) + \frac{1}{12n} + \mathcal{O}\left(n^{-2}\right) ,
\end{align}
the logarithm is
\begin{align}
    \log(\frac{D_A}{d_AD}) 
    =& \; - N \log(4) + \frac{N_B}{4} \log(\frac{N_B^4}{\left(\frac{N_B}{4} - m_1\right) \left(\frac{N_B}{4} - m_2\right) \left(\frac{N_B}{4} - m_3\right) \left(\frac{N_B}{4} - m_4\right)}) \nonumber \\ &
    +m_1 \log \left(\frac{N_B}{4}-m_1\right)+m_2 \log \left(\frac{N_B}{4}-m_2\right)+m_3 \log \left(\frac{N_B}{4}-m_3\right)+m_4 \log \left(\frac{N_B}{4}-m_4\right) \nonumber \\ &
    + \frac{1}{2}\log(\frac{N_B (\frac{N}{4})^4}{N 
    \left(\frac{N_B}{4} - m_1\right) \left(\frac{N_B}{4} - m_2\right) \left(\frac{N_B}{4} - m_3\right) \left(\frac{N_B}{4} - m_4\right)
    }) + \frac{5}{4N} + \frac{1}{12N_B} - \frac{1}{12 \left(\frac{N_B}{4}-m_1\right)} \nonumber \\ & -\frac{1}{12 \left(\frac{N_B}{4}-m_2\right)}  -\frac{1}{12 \left(\frac{N_B}{4}-m_3\right)} -\frac{1}{12 \left(\frac{N_B}{4}-m_4\right)} +  \mathcal{O}\left(N^{-3/2}\right).
\end{align}
We Taylor-approximate about $N = \infty$ and reorganize:
\begin{align}
    \log(\frac{D_A}{d_AD}) 
    &= - N_A \log(4)  + \frac{3}{2}\log(\frac{N}{N_B}) + \sum_i \left( - \frac{2m_i^2}{N_B} - \frac{8m_i^3}{3N_B^2} + \frac{4m_i^2}{N_B^2}  - \frac{16m_i^4}{3N_B^3}  \right)  + \frac{5}{4N} - \frac{5}{4N_B}    +   \mathcal{O}\left(N^{-3/2}\right). \label{eq:StrApprox_2}
\end{align}

The logarithm approximation~\eqref{app:Da/D_approx_2} and the ratio approximation~\eqref{eq:StrApprox_2} can now be substituted into the state-counting term~\eqref{app:eq_sub_in}. The summand varies slowly where its value is large, so we approximate the sum as an integral. Also, the integrand falls off quickly enough at large $|A_k|$ that we approximate the limits as $\pm \infty$. Evaluating the resulting Gaussian integrals, we obtain the commuting-charge state-counting term:
\begin{align}
     S(\expval{\rho_A}_{{\cal{C}}_0})  &= N_A \log(4)  - \frac{3}{2}\log(\frac{N}{N_B}) + \frac{3N_A}{N} + \frac{3 N_A}{4 N^2}-\frac{N_A^2}{2 N^2 N_B}  +   \mathcal{O}\left(N^{-3/2}\right).
\end{align}

\subsection{Noncommuting-charge model's state-counting term}\label{app:noncommuting_model}

The noncommuting charges share exactly one eigenspace, ${\cal{N}}_0$, specified as follows. Recall that the $a$ qubits' total-spin-squared operator, $\vec{S}_a^2$, has eigenvalues $s(s+1)$. Consider tensoring the $a$ qubits' $s = 0$ eigenspace onto the $b$ qubits' full Hilbert space. The product is the eigenvalue-0 eigenspace shared by $Q_{1,2,3}^\tot$. 

We calculate first the $a$ qubits' contribution to the state-counting term, then the $b$ qubits' contribution (App.~\ref{app:exact_2}). 
In App.~\ref{app:closed_3}, we approximate the state-counting term to order $\mathcal{O}(N^{-1})$, as is necessary for identifying differences from the commuting-charge model. 

\subsubsection{Exact expression for the noncommuting-charge model's state-counting term}\label{app:exact_2}

First, we calculate the $a$ qubits' contribution to the state-counting term. By the rules for angular-momentum addition, $s = \abs{s_A - s_B}, \abs{s_A - s_B} + 1, \hdots, s_A + s_B$. Therefore, $s=m=0$ only if $\abs{s_A - s_B} = 0$---equivalently, only if $s_A = s_B$ and $m_A=-m_B$.  This restriction constrains the global system to a subspace $\mathcal{N}_0$ of dimensionality
\begin{align}
    D &= C_{\frac{N}{2},\frac{N}{2}} = \frac{1}{\frac{N}{2}+1}{N \choose \frac{N}{2}}.
\end{align}

We now choose a basis for this subspace. A natural choice consists of states with quantum numbers $s_A=s_B$. If $s_A=s_B=0$, these basis states are tensor products. However, almost all the basis states correspond to $s_A=s_B>0$ and encode entanglement between $A$ and $B$, unlike the basis states chosen for the commuting-charge model. The noncommuting-charge basis states Schmidt-decompose as
\begin{align}
    \label{eq_Schmidt}
    |s_A,i,j\rangle = \sum_{m_A=-s_A}^{s_A} \frac{(-1)^{m_A}}{\sqrt{2s_A+1}} \,
    |s_A,m_A,i\rangle_A \,
    |s_B {=} s_A , m_B {=} -m_A, j\rangle_B .
\end{align}
The $i$ indexes the elements of an arbitrary orthonormal basis for the subsystem-$A$ subspace associated with the quantum numbers $s_A$ and $m_A$. This subspace is of dimensionality
\begin{align}
     d_A &= 
     C_{\frac{N_A}{2}+s_A,\frac{N_A}{2}-s_A} 
     = 
     \frac{2s_A+1}{\frac{ N_A}{2}+s_A+1} {N_A \choose \frac{N_A}{2} - s_A}~.
\end{align}
The $j$ in~\eqref{eq_Schmidt} indexes the elements of an arbitrary orthonormal basis for the subsystem-$B$ subspace associated with the quantum numbers $s_B$ and $m_B$. This subspace is of dimensionality
\begin{align}
     d_B &= 
     C_{\frac{N_B}{2}+s_B,\frac{N_B}{2}-s_B} 
     = 
     \frac{2s_B+1}{\frac{ N_B}{2}+s_B+1} {N_B \choose \frac{N_B}{2} - s_B}~.
\end{align}
The global system's maximally mixed state is
\begin{align}
\expval{\rho}_{{\cal{N}}_0} = \frac{1}{D} \sum_{s_A,i,j} \dyad{s_A,i,j}~.
\end{align} 
Tracing out subsystem $B$ yields
\begin{equation}
     \expval{\rho_A}_{{\cal{N}}_0} = \frac{1}{D} \sum_{s_A,m_A,i} \frac{d_B}{2s_A+1} \dyad{s_A,m_A,i}.
\end{equation}
Taking the spectral decomposition, we calculate the state's entropy and so the $a$ qubits' contribution to the state-counting term: 
\begin{align}
    S(\expval{\rho_A}_{{\cal{N}}_0})  = - \sum_{s_A = 0}^{\frac{N_A}{2}} & \; \frac{d_Ad_B}{D} \log(\frac{d_B}{D(2s_A+1)}) \label{app:eq_Da/D_ref_1}  \\
    = - \sum_{s_A=0}^{\frac{N_A}{2}} & \; \left(\frac{N}{2}+1\right) \frac{\left(\frac{N}{2}\right)!\left(\frac{N}{2}\right)!}{N!} \left(\frac{2s_A+1}{\frac{N_A}{2}+s_A+1} \right)
    \left( \frac{2s_A+1}{\frac{N_B}{2}+s_A+1} \right) {N_A \choose \frac{N_A}{2} - s_A} {N_B \choose \frac{N_B}{2} - s_A}  \nonumber \\ & \times
    \log\left( \frac{\left(\frac{N}{2}\right)!\left(\frac{N}{2}\right)!}{N!} \frac{(\frac{N}{2}+1)}{(\frac{N_B}{2}+s_A+1)} {N_B \choose \frac{N_B}{2} - s_A}\right). \label{eq:s_general}
\end{align}

We now calculate the $b$ qubits' contribution. $N_A$ unconstrained qubits have a state-counting term of $N_A\log(2)$. Adding $N_A\log(2)$ to Eq.~\eqref{eq:s_general} yields the noncommuting-charge state-counting term:
\begin{align}
    S(\expval{\rho_A}_{{\cal{N}}_0})   =  N_A\log(2) - \sum_{s_A=0}^{\frac{N_A}{2}} & \; \left(\frac{N}{2}+1\right) \frac{\left(\frac{N}{2}\right)!\left(\frac{N}{2}\right)!}{N!} \left(\frac{2s_A+1}{\frac{N_A}{2}+s_A+1} \right)
    \left( \frac{2s_A+1}{\frac{N_B}{2}+s_A+1} \right) {N_A \choose \frac{N_A}{2} - s_A} {N_B \choose \frac{N_B}{2} - s_A}  \nonumber \\ & \times
    \log\left( \frac{\left(\frac{N}{2}\right)!\left(\frac{N}{2}\right)!}{N!} \frac{(\frac{N}{2}+1)}{(\frac{N_B}{2}+s_A+1)} {N_B \choose \frac{N_B}{2} - s_A}\right).
\end{align}

\subsubsection{Closed-form approximation to the noncommuting-charge model's state-counting term} \label{app:closed_3}

First, we approximate the $\frac{d_Ad_B}{D}$ in Eq.~\eqref{app:eq_Da/D_ref_1} as a Gaussian function. We break $\frac{d_Ad_B}{D}$ into two factors, one consisting of factorials and the other of everything else:
$\frac{d_Ad_B}{D} = f(s_A) g(s_A)$, wherein
\begin{align}
    f(s_A) &\coloneqq  \frac{\left(\frac{N}{2}\right)!\left(\frac{N}{2}\right)!}{(N)!}\frac{(N_A)!}{\left(\frac{N_A}{2}+s_A\right)!\left(\frac{N_A}{2}-s_A\right)!}\frac{(N_B)!}{\left(\frac{N_B}{2}+s_B\right)!\left(\frac{N_B}{2}-s_B\right)!}\ \quad \text{and}\\
    \label{eq_g_s_A}
    g(s_A) &\coloneqq \left(\frac{2s_A+1}{\frac{N_A}{2}+s_A+1}\right) \left(\frac{2s_A+1}{\frac{N_B}{2}+s_A+1}\right)\left(\frac{N}{2}+1\right).
\end{align}
We Taylor-expand $\log\LParen f(s_A) \RParen$ around its maximum, $s_A = 0$, to $\mathcal{O}\left(N^{-1}\right)$, assuming $s_A^2\sim N$. Then, we exponentiate the result:
\begin{align}
   f(s_A) &= \sqrt{\frac{2N}{N_AN_B\pi}} \exp(\frac{-2N s_A^2}{N_AN_B})\left[1 + \frac{1}{4N} - \frac{1}{4N_A} - \frac{1}{4N_B} + \frac{2s_A^2}{N_A^2} + \frac{2s_A^2}{N_B^2}  -\frac{4s_A^4}{3N_A^3} -\frac{4s_A^4}{3N_B^3} + \mathcal{O}\left(N^{-2}\right)\right]. \label{eq:s0_fsA}
\end{align}
Next, we expand $g(s_A)$ [Eq.~\eqref{eq_g_s_A}]:
\begin{align}
  g(s_A) =& \; \frac{8Ns_A^2}{N_AN_B}\left[1
  +\frac{1}{s_A}
  -\frac{2 s_A}{N_A}
  -\frac{2 s_A}{N_B}
  +\frac{2}{N}
  -\frac{4}{N_A}
  -\frac{4}{N_B}
  +\frac{1}{4 s_A^2}
  +\frac{4 s_A^2}{N_A^2}
  +\frac{4 s_A^2}{N_B^2}
  +\frac{4 s_A^2}{N_A N_B}
  + \mathcal{O}\left(N^{-3/2}\right)\right]. \label{eq:s0_gsA}
\end{align}
The right-hand sides of~\eqref{eq:s0_fsA} and~\eqref{eq:s0_gsA} multiply to
\begin{align}
    \frac{d_Ad_B}{D} 
    =& \; \frac{4(2N)^{\frac{3}{2}}}{(N_AN_B)^{\frac{3}{2}}\sqrt{\pi}} \, s_A^2
    \exp(\frac{-2N s_A^2}{N_AN_B})\bigg[
    1+ \frac{1}{s_A} - \frac{2Ns_A}{N_AN_B} + 
    \frac{9}{4N} - \frac{17N}{4N_AN_B} + \frac{6s_A^2}{N_A^2} + \frac{6s_A^2}{N_B^2} + \frac{4s_A^2}{N_AN_B}  \nonumber \\ &+ \frac{1}{4s_A^2}  - \frac{4s_A^4}{3N_A^3} -\frac{4s_A^4}{3N_B^3} +  \mathcal{O}\left(N^{-3/2}\right)\bigg]. \label{app:Da/D_approx_1}
\end{align}
We check that this function is normalized to $O( N^{-3/2 } )$ (as $\frac{d_Ad_B}{D}$ must be normalized), but omit the details of the check.

Having approximated the first factor in the state-counting term~\eqref{app:eq_Da/D_ref_1}, we proceed to the second.
According to the Stirling approximation~\eqref{eq_Stirling},
the logarithm is
\begin{align}
    \log(\frac{d_B}{D(2s_A+1)})
    =& \; - N \log (2) + \frac{N_B}{2} \log \left(\frac{N_B^2}{\left(\frac{N_B}{2}-s_A\right) \left(\frac{N_B}{2}+s_A\right)}\right)
    +s_A \log \left(\frac{\frac{N_B}{2}-s_A}{\frac{N_B}{2}+s_A}\right) 
    \nonumber \\ &
    + \frac{1}{2} \log \left(\frac{NN_B}{4\left(\frac{N_B}{2}-s_A\right) \left(\frac{N_B}{2}+s_A\right)}\right) 
    + \log \left(\frac{\frac{N}{2}+1}{\frac{N_B}{2}+s_A+1}\right)
    \nonumber \\ &
    + \frac{1}{4N} + \frac{1}{12N_B}  - \frac{1}{12\left(\frac{N_B}{2}-s_A\right)} - \frac{1}{12\left(\frac{N_B}{2}+s_A\right)} + \mathcal{O}\left(N^{-2}\right).
\end{align}
Taylor-approximating about $N = \infty$ yields
\begin{align}
     \log(\frac{d_B}{D(2s_A+1)})&= - N_A \log (2) + \frac{3}{2}\log(\frac{N}{N_B})  - \frac{2s_A^2}{N_B} - \frac{2s_A}{N_B}
     - \frac{4s_A^4}{3N_B^3} + \frac{4s_A^2}{N_B^2}
       + \frac{9}{4N} -\frac{9}{4N_B}  +  \mathcal{O}\left(N^{-3/2}\right). \label{eq:StrApprox_1}
\end{align}

We can now substitute the logarithm~\eqref{eq:StrApprox_1} and
the $d_Ad_B/D$ factor~\eqref{app:Da/D_approx_1} into the state-counting term~\eqref{app:eq_Da/D_ref_1}. Since the summand varies slowly where its value is large, we approximate the sum as an integral. Also, since the integrand falls off rapidly at large $s_A$, we approximate the integral's upper limit with $\infty$. Evaluating the integral, we calculate the $a$ qubits' contribution to the state-counting term. Adding the $b$ qubits' state-counting term, $N_A\log(2)$, we obtain the noncommuting-charge state-counting term:
\begin{equation}
    \expval{S_{\Ent}}_{\cal{S}}  = N_A \log (4) - \frac{3}{2}\log(\frac{N}{N_B}) + \frac{3N_A}{2N} +\frac{3 N_A}{4 N^2} + \frac{N_A^2}{2 N^2 N_B}  +  \mathcal{O}\left(N^{-3/2}\right).
\end{equation}

\section{How our models' charges restrict the microcanonical subspaces}

The main text posits an explanation for why, in the microcanonical-subspace study, the noncommuting-charge Page curve lies above the commuting-charge Page curve. We propose another explanation, using specifics of our models, here. To what extent this reasoning generalizes beyond those models merits further study.

Consider beginning with an unconstrained system, then restricting the Hilbert space to the eigenvalue-0 eigenspace of $Q_1^\tot$, then restricting further to the eigenvalue-0 eigenspace of $Q_2^\tot$, then restricting to the eigenvalue-0 eigenspace of $Q_3^\tot$. The first two restrictions already restrict the system to the $s=0$ subspace; the third restriction is redundant. 

Now, consider undertaking the same process but replacing the $Q_\alpha^\tot$'s with $C_\alpha^\tot$'s. The first two restrictions only partially imply the third, which therefore constrains the Hilbert space nontrivially. (Appendix~\ref{app_Partial_C_Constraints} contains a proof.)
One might therefore expect the microcanonical subspace to be larger when defined by our three noncommuting charges than when defined by our three commuting charges. We have confirmed this expectation by direct calculation. Furthermore, the available Hilbert space's dimensionality upper-bounds the entanglement entropy [Eq.~\eqref{eq:EE_onestate}]. Hence the noncommuting charges should enable more entanglement---a higher Page curve---than the commuting charges do.

\subsection{Constraining $C_1^\tot$ and $C_2^\tot$ constrains $C_3^\tot$ only partially}
\label{app_Partial_C_Constraints}

Consider an unconstrained system of $N$ 4-level qudits. Consider restricting the Hilbert space to the eigenvalue-0 eigenspace of $C_1^\tot$, then restricting further to the eigenvalue-0 eigenspace of $C_2^\tot$, and then restricting to the eigenvalue-0 eigenspace of $C_3^\tot$. The first two restrictions partially imply the third, which constrains the Hilbert space nontrivially. We prove this claim here.

The local charges $C_{1,2,3}$ share four eigenstates,
the maximally entangled Bell states~\cite{nielsen2002quantum}.
They are, if $\ket{\uparrow}$ and $\ket{\downarrow}$ denote the $Z$ eigenstates,
\begin{align}
    & \ket{\mathcal{B}_1} \coloneqq \frac{1}{\sqrt{2}}\left(
    \ket{\downarrow}_a\ket{\uparrow}_b - \ket{\uparrow}_a\ket{\downarrow}_b \right), \quad
    \ket{\mathcal{B}_2} \coloneqq \frac{1}{\sqrt{2}}\left(
    \ket{\downarrow}_a\ket{\downarrow}_b - \ket{\uparrow}_a\ket{\uparrow}_b \right), \\
    & \ket{\mathcal{B}_3} \coloneqq \frac{1}{\sqrt{2}}\left(
    \ket{\downarrow}_a\ket{\downarrow}_b + \ket{\uparrow}_a\ket{\uparrow}_b \right), 
    \quad \text{and} \quad
    \ket{\mathcal{B}_4} \coloneqq \frac{1}{\sqrt{2}}\left(
    \ket{\downarrow}_a\ket{\uparrow}_b + \ket{\uparrow}_a\ket{\downarrow}_b \right).
\end{align}
Denote by $\rho_j$ the $j^\th$ qubit's reduced state,
which has a weight $\bra{ \mathcal{B}_k } \rho_j \ket{ \mathcal{B}_k }$ on the $k^\th$ Bell state. Summing over qudits yields the total population $P_k \coloneqq \sum_{j=1}^N \bra{ \mathcal{B}_k } \rho_j \ket{ \mathcal{B}_k }$.

If the system is in an eigenvalue-0 eigenstate of $C_{1}^{\tot}$, then $P_1 + P_2 = P_3 + P_4$. 
If the system is in an eigenvalue-0 eigenstate of $C_{2}^{\tot}$, then $P_1 + P_3 = P_2 + P_4$. Together, these constraints imply $P_1 = P_4$ and $P_2 = P_3$. Furthermore, 
$\expval{ C_{3}^{\tot} } = P_2 + P_3 - P_1 - P_4$. This expectation value, under the $C_{1,2}^{\tot}$ constraints, is restricted to $2(P_2 - P_1)$, which need not vanish. 
Thus, the first two charges do not restrict the $C_{3}^{\tot}$ expectation value completely. 
Contrarily, if in an eigenstate of $Q_{1,2}^\tot$, the system is in the eigenvalue-0 eigenstate of $Q_3^\tot$. Hence $C_{1,2}^\tot$ restrict the Hilbert space less than $Q_{1,2}^\tot$ do.

\new{
\section{How sequentially introduced charges change the Page curve: superadditively, subadditively, or additively}\label{app:additivity_ansatz}
}

Figure~\ref{fig:s0_change_EE} shows Page curves constructed from microcanonical subspaces. At finite $N$, the curves violate an expectation that one might gather from earlier literature. We explain the expectation, discuss the violation, and provide numerical evidence for the expectation in the thermodynamic limit (as $N \to \infty$).

Consider beginning with an unconstrained $N$-site system, restricting the Hilbert space to the eigenvalue-0 eigenspace of $C_1^\tot$, then restricting further to the eigenvalue-0 eigenspace of $C_2^\tot$, and then restricting to the eigenvalue-0 eigenspace of $C_3^\tot$.
One might expect that, as more charges were introduced, each successive charge would lower the Page curve by the same amount as the last charge. Such lowering has been argued to happen in the thermodynamic limit, with commuting charges~\cite{altland2022maximum}. We call an expectation of such lowering the \textit{additivity ansatz}. One might posit it, expanding on~\cite{altland2022maximum}, (i) for noncommuting charges in the thermodynamic limit and (ii) for commuting and noncommuting charges at finite $N$.

If the additivity ansatz were true, the Page curve (for three equivalent commuting or noncommuting charges) could be constructed as follows. Consider restricting the global Hilbert space to one charge's eigenvalue-0 eigenspace (any $C_\alpha^\tot$ or $Q_\alpha^\tot$---which one does not affect the curve). The corresponding Page curve, we denote by $\expval{S_{\Ent}}_{\cal{S}}^{(1)}$. Recall that
$ \expval{S_E}_{\cal{H}}$ denotes the unrestricted Page curve. The additivity ansatz predicts the Page curve
$ \expval{S_E}_{\cal{H}} - 3\left( \expval{S_E}_{\cal{H}} -  \expval{S_{\Ent}}_{\cal{S}}^{(1)}\right)$ for our models with three equivalent charges constrained in each.

\begin{figure}
     \centering
         \includegraphics[width=0.49\textwidth]{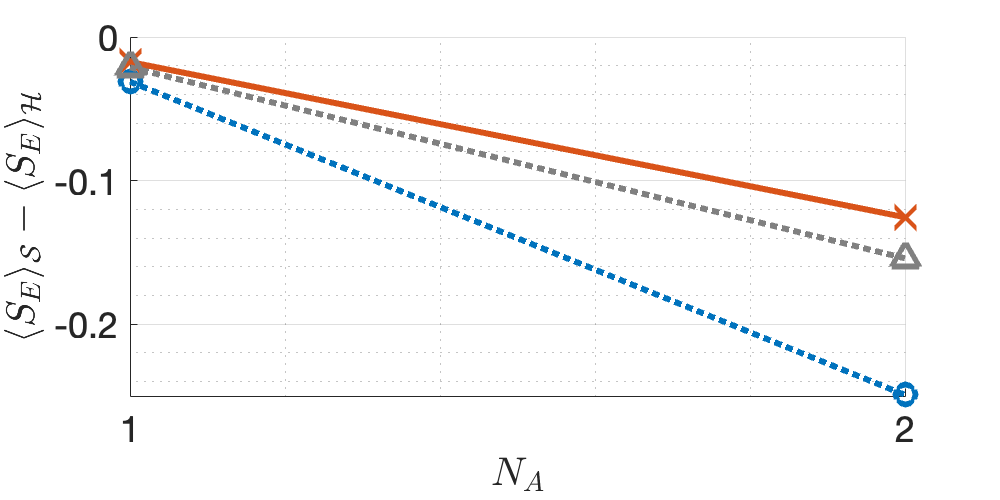}
         \includegraphics[width=0.49\textwidth]{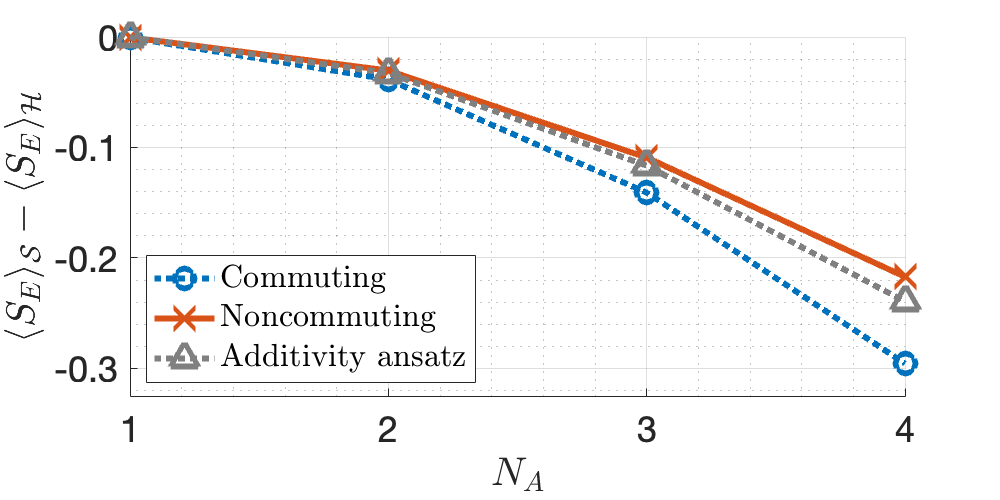}
        \caption{\caphead{Testing the additivity ansatz.} 
        $\expval{S_{\Ent}}_{\cal{S}}$ denote any Page curve restricted by charges; and $ \expval{S_E}_{\cal{H}}$, the unrestricted Page curve. The red x's form the noncommuting-charge model's Page curve, and the circular blue markers form the commuting-charge model's Page curve. Both curves were calculated using microcanonical subspaces. The gray triangles illustrate the additivity ansatz.
        }
    \label{fig:add_ansatz}
\end{figure}

Figure~\ref{fig:add_ansatz} tests this prediction at finite $N$. The gray triangles form the additivity-ansatz curve. It lies below the noncommuting-charge Page curve (red x's), which are therefore superadditive. The ansatz curve also lies above the commuting-charge Page curve (blue circles), which are subadditive. Hence the additivity ansatz breaks in a commutation-dependent manner at finite $N$. However, all three curves converge as $N$ grows. We hence provide numerical evidence for the additivity ansatz, supported analytically in~\cite{altland2022maximum} and in App.~\ref{app:analytical} above, in the thermodynamic limit.

\section{Analogous approximate microcanonical subspaces}\label{app:eg_amc_probs}

The main text specifies how to construct AMC subspaces in the noncommuting-charge model. We augment this explanation with examples. Then, we explain how to construct analogous AMC subspaces in the commuting-charge model. We also specify the six analogous-AMC-subspace pairs reported in the main text.

First, we review how to construct AMC subspaces in the noncommuting-charge model. Denote by $2m$ the $Z_a^\tot$ eigenvalue. $Z_a^\tot$ shares eigenstates with $\vec{S}_a^2$. Shared eigenstates labeled by the same two quantum numbers form the \emph{$(s, m)$ eigenspace}. Some such eigenspaces are AMC subspaces, we find by direct calculation. For each $(s, m)$ value, we calculate the probability distributions $p_\alpha^{\cal{N}}(\gamma)$. Each distribution should exhibit one peak for the eigenspace to satisfy the AMC subspace's definition. 
$p_3^{\cal{N}}(\gamma)$, being a Kronecker delta function in the $(s, m)$ subspace, meets this criterion. Also, according to direct calculation, $p_1^{\cal{N}}(\gamma) =  p_2^{\cal{N}}(\gamma)$ for all $\gamma$. Hence we need calculate only $p_1^{\cal{N}}(\gamma)$ to check whether an $(s, m)$ eigenspace is an AMC subspace. Table \ref{tab:AMC_NC_probability} presents these distributions for $s \leq 4$. Whenever $s=m$, each distribution exhibits one peak. Therefore, each $(s, m{=}s)$ subspace qualifies as an AMC subspace.

\begin{table}[]
 \centering
\begin{tabular}{@{}cccccccccc@{}}
\toprule
\multirow{2}{*}{$(s,m)$} & \multicolumn{9}{c}{Possible measurement outcomes} \\ 
 & -4 & -3 & -2 & -1 & 0 & 1 & 2 & 3 & 4 \\
\midrule
$(1,0)$ &   &  &  & 0.500 & 0 & 0.500 &  &  &  \\ 
$(1,1)$ &   &  &  & 0.250 & 0.500 & 0.250 &  &  &  \\ 
$(2,0)$ &   &  & 0.375 & 0 & 0.250 & 0 & 0.375 &  &  \\ 
$(2,1)$ &   &  & 0.250 & 0.250 & 0 & 0.250 & 0.250 &  &  \\ 
$(2,2)$ &   &  & 0.063 & 0.250 & 0.375 & 0.250 & 0.063 &  &  \\ 
$(3,0)$ &   & 0.313 & 0 & 0.188 & 0 & 0.188 & 0 & 0.313 &  \\ 
$(3,1)$ &   & 0.234 & 0.156 & 0.016 & 0.188 & 0.016 & 0.156 & 0.234 &  \\ 
$(3,2)$ &   & 0.094 & 0.250 & 0.156 & 0 & 0.156 & 0.250 & 0.094 &  \\ 
$(3,3)$ &   & 0.016 & 0.094 & 0.234 & 0.313 & 0.234 & 0.094 & 0.016 &  \\ 
$(4,0)$ &  0.273 & 0 & 0.156 & 0 & 0.141 & 0 & 0.156 & 0 & 0.273 \\
$(4,1)$ &  0.219 & 0.109 & 0.031 & 0.141 & 0 & 0.141 & 0.031 & 0.109 & 0.219 \\ 
$(4,2)$ & 0.109 & 0.219 & 0.063 & 0.031 & 0.156 & 0.031 & 0.063 & 0.219 & 0.109 \\
$(4,3)$ & 0.031  & 0.141 & 0.219 & 0.109 & 0 & 0.109 & 0.219 & 0.141 & 0.031 \\ 
$(4,4)$ & 0.004  & 0.031 & 0.109 & 0.219 & 0.273 & 0.219 & 0.109 & 0.031 & 0.004 \\ \bottomrule
\end{tabular}
    \caption{\caphead{Probabilities $p_1^{\cal{N}}(\gamma)$ that characterize $(s,m)$ eigenspaces.}
    Denote by $\ket{\psi}$ any state from an $(s,m)$ eigenspace of the noncommuting-charge model. Measuring $Q_{1}^\tot$ yields outcome $\gamma$ with some probability. This probability, averaged over the $\ket{\psi}$, we denote by $p_{1}^{\cal{N}}(\gamma)$. The possible measurement outcomes range from $-s$ to $s$. The probabilities $p_{1}^{\cal{N}}(\gamma)$ are listed for each $(s,m)$ and are independent of the system size, $N$. $p_{1}^{\cal{N}}(\gamma)$ has exactly one peak only if $s=m$.}
    \label{tab:AMC_NC_probability}
\end{table}

Having identified AMC subspaces defined by noncommuting charges, we construct analogs defined by commuting charges. For each $N$, we identify the eigenspaces shared by $C_{1,2,3}^\tot$. For consistency with the noncommuting-charge model, we keep only the eigenvalue-$m$ eigenspaces of $C_3^{\tot}$. For each shared eigenspace, we calculate the distributions $p_\alpha^{\cal{C}}(\gamma)$.
If they equal their noncommuting-charge counterparts $p_\alpha^{\cal{N}}(\gamma)$ (criterion~\ref{crit5}), the eigenspace forms an analogous AMC subspace. 

An illustrative example is parameterized by $N=8$ and (in the noncommuting-charge model) $s=m=1$. We keep only the eigenvalue-$1$ eigenspaces of $C_3^{\tot}$. Denote by $c_x$ the $C_1^\tot$ eigenvalues and by $c_y$ the $C_2^\tot$ eigenvalues. 
We label by $(c_x, c_y, 1)$ the eigenspaces shared by $C_{1,2,3}^\tot$. For consistency with the noncommuting-charge model, we ignore any eigenspaces in which $c_x > s$ or $c_y > s$. Four eigenspaces remain: $(0, -1, 1)$, $(-1, 0, 1)$, $(1, 0, 1)$, and $(0, 1, 1)$. Each is of dimensionality 1680.
The candidate AMC subspace is the union of these four subspaces and is of dimensionality 6720. 
These dimensionalities fix the probabilities $p_{1}^{\cal{C}}(\gamma)$. For example, $p_{1}^{\cal{C}}(0) = (1680 \times 2)/6720 = 0.5$. The remaining probabilities are $p_{1}^{\cal{C}}(-1) = 0.25$ and $p_{1}^{\cal{C}}(1) = 0.25$.
This distribution equals the corresponding $p_1^{\cal{N}}(\gamma)$.
Checking every eigenvalue-$m$ eigenspace of $C_3^{\tot}$, we find six eigenspaces for which $p_\alpha^{\cal{C}}(\gamma) = p_\alpha^{\cal{N}}(\gamma) \; \forall \alpha, \gamma$, satisfying criterion~\ref{crit5}.

We have identified six pairs of parallel (commuting-charge and noncommuting-charge) AMC subspaces. The pairs are labeled by $s=m=1,N/2$ and $N=4,8$, as well as by $s=m=N/2$ and $N=2,6$. (Computational limitations restrict us to $N \leq 8$.) Table \ref{tbl:AMC_comparison} compares the two Page curves formed from each subspace pair. We compare the curves at their midpoints, $N_A = N/2$. The percent difference between the two curves varies from $0.199 \%$ to $3.06 \%$ across the subspace pairs. Hence noncommuting charges increase the average entanglement entropy in AMC subspaces as in microcanonical subspaces.

\begin{table}[]
\setlength{\tabcolsep}{0.5em}
\def\arraystretch{1.25}
\begin{tabular}{@{}cccccc@{}}
\toprule
$N$ & $s=m$ & NC & C & NC $-$ C& $\%$ diff.  \\ \midrule
4 & 1 & $-0.455$ & $-0.479$ & 0.024 & 5.112  \\
8 & 1 & $-0.364$ & $-0.390$ & 0.027 & 7.106 \\
2 & $N/2$ & $-0.587$ & $-0.589$ & 0.002 & 0.362\\ 
4 & $N/2$ & $-1.350$ & $-1.354$ & 0.004 & 0.272 \\
6 & $N/2$ & $-2.074$ & $-2.086$ & 0.012 & 0.600  \\
8 & $N/2$ & $-2.770$ & $-2.788$ & 0.017 & 0.625 \\ \bottomrule 
\end{tabular} 
\caption{\caphead{Differences between Page curves, constructed from approximate microcanonical subspaces, at $N_A = N/2$.} The Page curves' values at $N_A = N/2$ are listed for various $N$ and $s=m$ values. We abbreviate ``difference'' with ``diff.,'' ``noncommuting'' with ``NC,'' and ``commuting'' with ``C.''
} \label{tbl:AMC_comparison}
\end{table}

\end{appendices}

\bibliography{apssamp}
\end{document}